# Electron acceleration in laser turbulence

M. Sherlock

*Lawrence Livermore National Laboratory, California 94551, United States.*

(Dated: September 30, 2020)

We demonstrate that electrons can be efficiently accelerated to high energy in spatially non-uniform, intense laser fields. Laser non-uniformities occur when a perfect plane wave reflects off a randomly perturbed surface. By solving for three-dimensional particle trajectories in the electromagnetic field of a randomly perturbed laser, we are able to generate electron energy spectra with temperatures well above the ponderomotive scaling, as observed in many experiments. The simulations show that high electron energies can be achieved by the laser fields alone, without the need for plasma fields. The characteristic temperatures of the electron spectra are determined by the characteristic features of the laser field turbulence. The process is very rapid, occuring on $10-50fs$ timescales, indicating it is likely a dominant acceleration mechanism in short-pulse laser-solid interactions when the intensity is at or below the relativistic threshold. A simple analytic model shows how electrons can reach high energy by undergoing repeated acceleration in laser wavelets for short periods of time.

**PACS numbers:** PACS numbers.

Intense, short-pulse laser-plasma experiments allow us to readily produce extreme states of matter in the laboratory. Many laboratories around the world, at both the university and national scale, are readily able to produce extreme states of matter by directing very intense, short, laser pulses onto specially designed targets. This allows us to study a wide variety of physical processes, from particle acceleration and intense radiation sources to laboratory astrophysics and positron generation. The possibility of observing high-field phenomena, such as quantum electrodynamics processes [1, 2], or of utilizing these formidable tools for igniting dense fusion plasmas [9], are active areas of research. The theoretical study of the non-linear and multiple-scale behavior of particles and matter under these conditions has been greatly aided by the particle-in-cell simulation (PIC) technique [3]. Large-scale, multi-dimensional simulations are themselves often so complex that isolating and clearly identifying the dominant physical processes can be challenging, and arriving at unambiguous conclusions is difficult.

One striking feature of intense laser-plasma experiments is the observation of thermal electron energy spectra, i.e. spectra with functional form close to $dN/dE \sim e^{-E/T_e}$ (E is the electron energy and $T_e$ the best-fit temperature). The thermal form of the spectra is strongly suggestive of a stochastic acceleration process, and many such processes have been proposed, mostly on the basis of one-dimensional simulations [10, 12–14]. The most well known heating mechanism is the ponderomotive acceleration, which predicts electrons gain kinetic energy equal to the ponderomotive potential $\varphi_p = \left(\sqrt{1+I\lambda_{\mu m}^2/1.37\times 10^{18}} - 1\right)mc^2$ (with intensity $I$ in $Wcm^{-2}$, wavelength $\lambda_{\mu m}$ in $\mu m$) by undergoing a single $\boldsymbol{j}\times\boldsymbol{B}$ oscillation in the wave close to the critical surface. After crossing the critical surface, electrons retain this energy because they are no longer in the vicinity of strong fields. Robinson [4] has shown one route to electron energy gain is the "breaking of adiabaticity" by some force of non-laser origin (i.e. not the Lorentz force associated with the laser plane wave fields). Examples of non-laser forces are the electric field formed in ion channels [4], the electrostatic sheath field [13] and plasma waves [5]. Other processes have been identified which occur in two [15] and three dimensions [6]. These mechanisms are associated with electron motion across large-scale features (such as the ion channel, plasma sheath, laser spot or target dimensions), and operate on relatively long timescales because the anti-dephasing is driven by plasma fields, which are relatively weak in comparison to the laser fields. In contrast, the mechanism we describe in this letter occurs on much shorter timescales, and is therefore likely a dominant acceleration mechanism. For example, by the process described here, electrons can be accelerated to multi-MeV energies within $10fs$ at the non-relativistic intensity of $I = 4\times 10^{17}Wcm^{-2}$. The mechanism is based on electrons accelerated by non-uniformities in the laser fields themselves (which are approximately an order of magnitude stronger than the induced plasma fields). We show that the random nature of typical laser non-uniformities ("laser turbulence") causes electrons to undergo stochastic motion and non-adiabatic acceleration, producing electron energy spectra with characteristic energy determined by the length of time electrons spend in the turbulence, as well as the spectral content of the turbulence. In short scale length plasmas this produces the well-known ponderomotive temperature scaling observed in early PIC simulations [7], in longer scale-length plasmas this can produce temperatures more than an order of magnitude above ponderomotive, as observed in many experiments. The heating mechanism can be identified unambiguously because no plasma fields are necessary. A simple analytic model complements the simulations by showing how electrons cumulatively gain energy in a series of "kicks" by staying in phase with the wave for short periods of time.

Since the earliest two-dimensional PIC simulations, irregular, non-plane-wave spatial structure in the laser fields has been noted [7, 8, 20]. These non-uniformities are referred to as "disturbances", "ripples", "filaments" or "fluctuations". Brady [8] studied this phenomenon in depth, showing that for intensities above $I \approx 5 \times 10^{17} W cm^{-2}$ the non-uniformities, which have spatial scale on the order of the laser wavelength ($\lambda_0$) and localized intensity maxima $\approx 4\times$ the incident intensity, are seeded by the Raman instability. Laser envelope and phase distortions also frequenctly occur as a result of imperfections in the laser system itself, or in the form of speckles. Whatever the cause, it is clear that when a uniform plane-wave reflects off a perturbed surface, the reflected light will contain significant non-uniformities if the surface contains depth perturbations on the order of $\lambda_0$, as seen in multi-dimensional PIC simulations. These diffractive non-uniformities, or fluctuations, have spatial scale $\gtrsim \lambda_0$, sometimes with a longer length scale in the direction along the laser k-vector (z), depending on the nature of the surface perturbations and distance from the surface. Filamented light is also sometimes associated with density filaments rather than diffraction, although these filaments do not appear to be a dominant cause of absorption, and electrons do not become trapped inside such filaments [18]. We use the term *fluctuations* to distinguish the time-dependent non-uniformities discussed here from their time-independent counterpart (*speckles*), although both phenomena arise from diffraction.

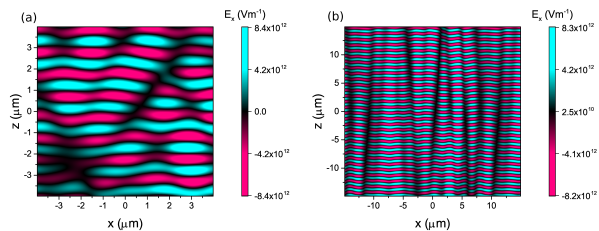

Figure 1. The transverse electric field for the two cases considered in the text: (a) short case with $L_z = 1.2\mu m$ and (b) long case with $L_z = 12.2\mu m$. The field is shown at an arbitrary time and the average intensity is $10^{18}Wcm^{-2}$. Note that the two figures are not at the same scale.

To study the acceleration of electrons in laser turbulence, we have developed the *Quartz* simulation code, which specifies the laser fields analytically, thereby removing numerical heating associated with the computational grid, which can be a source of significant error when studying a heating process. In order to obtain unambiguous results, the plasma response is ignored, so that all effects can be attributed to the laser fields only. This has the added advantage of making 3D simulations computationally feasible. The electron equations of motion ($d\mathbf{p}/dt = -e(\mathbf{E} + \mathbf{v} \times \mathbf{B})$ and $d\mathbf{x}/dt = \mathbf{v}$) are updated numerically with 4th-order relativistic Runge-Kutta integration, which provides excellent energy conservation ($\mathbf{p}$ is the momentum, $\mathbf{v}$ is the velocity, $\mathbf{E}$ is the electric field, $\mathbf{B}$ is the magnetic field, e is the electron charge and m its mass). The fields of the reflected wave are obtained from the vector potential $\boldsymbol{A}(\boldsymbol{x},t) = \boldsymbol{A}_R/2N \sum_{j,l=-N}^{N-1} \exp\left\{i\left(-\pi\frac{jx+ly}{NL_\perp} - 2\frac{j^2+l^2}{N^2}\frac{z}{L_z} + \phi_{j,l}\right)\right\}$, where $\mathbf{k} = k_z\hat{\mathbf{z}}$ is the laser wavevector, $\omega$ the laser frequency, $2N$ is the number of surface perturbations considered per dimension (typically $2N = 18$), $R_\perp$ is the generated mean transverse intensity radius, $L_z$ the generated mean longitudinal intensity length, $\mathbf{A}_R = \hat{\mathbf{x}}A_{0R}\exp[i(-\mathbf{k}.\mathbf{x} - \omega t)]$, $A_{0R}$ the amplitude of the vector potential of the reflected wave, and $\phi_{j,l}$ are phase factors randomly generated with value 0 or $\pi$. The total potential is the sum of the reflected wave and the incident wave, $\mathbf{A}_I = \hat{\mathbf{x}}A_{0I}\exp[i(\mathbf{k}.\mathbf{x} - \omega t)]$, corresponding to p-polarization at normal incidence. In this study, the reflected wave intensity is assumed to be 0.8 that of the incident wave, representing an absorption fraction of $\alpha = 0.2$, typical of short-pulse laser-matter interactions ($1 - \alpha = A_{0R}^2/A_{0I}^2$). The above expression for $\boldsymbol{A}$ is the vector potential envelope obtained using Fourier optics for a speckled laser beam [19]. One could envisage calculating the fields from the Kirchoff integral if the surface is specified, however this would necessitate a means of specifying the surface analytically. The advantage of the above approach is that the fluctuation turbulence is generated with Gaussian statistics whose mean properties can be specified by the two parameters $R_\perp$ and $L_z$. This enables, for example, a spectrum of mostly "short" ($R_\perp \approx L_z \approx \lambda_0$) or mostly "long" ($L_z \gg R_\perp \approx \lambda_0$) fluctuations to be studied independently. Real systems with long scale-length plasmas are likely to contain both short and long features, while those with short scale-length plasmas will mostly contain short features. In Fig. 1 we plot the instantaneous laser electric field for two cases: (a) $R_\perp = 1\mu m$ and $L_z = 1.2\mu m$ (referred to as the "short case"), (b) $R_\perp = 1\mu m$ and $L_z = 12.2\mu m$ (referred to as the "long case"), on an arbitrary x-z plane at arbitrary time. The fields $\boldsymbol{A}(\boldsymbol{x},t)$ are infinitely periodic in each direction, so that boundary effects do not complicate the analysis.

In this letter we study the acceleration of electrons in fields with short and long scale non-uniformities and obtain a scaling relation for slope temperature as a function of intensity. We first consider in detail the spectra for the short case for a relatively low average incident intensity of $I = 4 \times 10^{17} W cm^{-2}$, in Fig.2. This intensity is of particular interest because according to ponderomotive scaling [7], the slope temperature should be non-relativistic ($\varphi_p \approx 70 keV$), since the electron quiver energy in the wave is non-relativistic. However, some experiments [16–18, 22] in this intensity range have measured temperatures an order of magnitude above ponderomotive. In our simulations, electrons very rapidly reach a slope temperature of $T_e \approx 210 keV$ in the first 10fs, then proceed to increase in temperature less rapidly, reaching $T_e \approx 240 keV$ by 50fs. In the case of long fluctuations (Fig. 2), electrons gain energy more rapidly, reaching a slope temperature of $T_e \approx 0.5 MeV$ in 10fs, then proceeding to generate spectra that deviate from a simple thermal form (similar to the non-thermal spectra observed in [16, 22, 23]). To estimate the effective temperature for non-thermal spectra, we introduce the parameter $E_{tail}$ defined as the average energy of electrons in the upper 90% of energy ranges - i.e. the average energy of all electrons with energy $E > E_{max}$, where $E_{max}$ is the maximum particle energy. This provides a convenient means of excluding the low energy bulk, which accelerates slowly and is unlikely to reach a detector, and gives a result which is approximately independent of the number of computational particles used. For the long case, we find $E_{tail} \approx 1.1 MeV$ at 10fs, rising to $E_{90} \approx 3.3 MeV$ by 50fs, which is remarkably high given the relatively low average intensity.

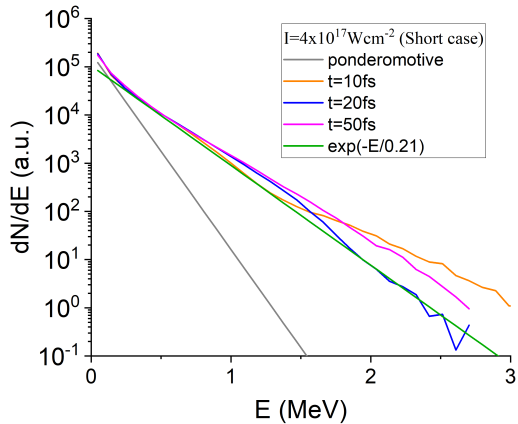
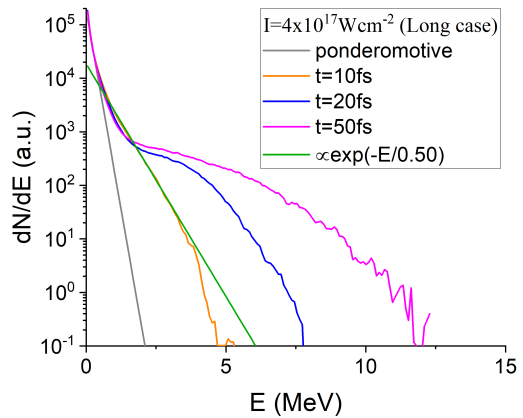

Figure 2. The electron spectrum at various times for the short case with average intensity $4 \times 10^{17} Wcm^{-2}$. Also shown is an exponential fit with slope $T_e = 0.21 MeV$.

Figure 3. The electron spectrum at various times for the long case with average intensity=4e17. Also shown is an exponential fit with slope $T_e = 0.5 MeV$.

This mechanism may explain the thermal nature of ponderomotive scaling. Although ponderomotive scaling [7] is widely cited, there is currently no theoretical understanding of the process that creates the associated quasi-thermal spectrum: according to the basic principle, electrons gain kinetic energy equal to the ponderomotive potential $\varphi_p = \left(\sqrt{1 + I\lambda_{\mu m}^2/1.37 \times 10^{18}} - 1\right) mc^2$ (with intensity $I$ in $Wcm^{-2}$, wavelength $\lambda_{\mu m}$ in $\mu m$) by undergoing a single oscillation in the wave close to the critical surface. After crossing the critical surface, electrons retain this energy because they are no longer in the vicinity of strong fields. However, the original simulations [7] showed thermal spectra, indicating electrons experience a wide range of energy gains, with sub-ponderomotive being the most probable and energy gains $\gg \varphi_p$ less probable (but nonetheless observed).

We summarize a range of simulations in Fig. 3, where we plot $E_{tail}$ (at 10fs) as a function of intensity for both the short and long cases, along with ponderomotive scaling. Note that the short and long cases merge to a common energy in the relativistic intensity range. The experimental data at low intensity are taken from [17, 18] and [16] at high intensity.

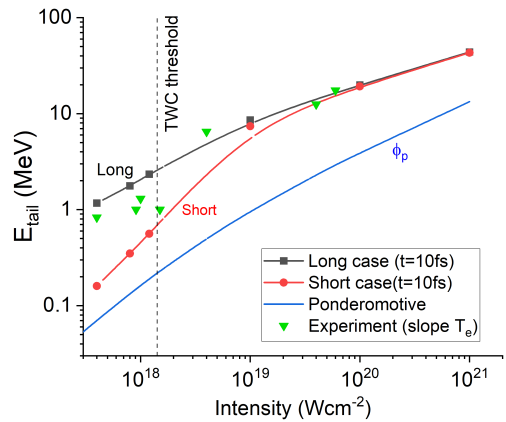

Figure 4. The average electron energy in the tail ($E_{tail}$) as a function of intensity. The vertical line marks the approximate intensity above which two-wave chaos occurs. Experimental data are shown in green. Solid curves are b-spline fits to the data.

We have developed a simple Chirikov map which reproduces the essential features of the particle simulations and gives analytic insight into the acceleration mechanism. This is based on the simple concept of dividing space into a collection of relatively small plane waves ("wavelets") separated by field nulls where the phase changes abruptly. Each wavelet is an isolated plane wave of finite length $L_z$ (along the direction of laser propagation) and constant intensity $\mathbf{A}_0$, with unique phase ($\phi_0$) and field $\mathbf{A}(z,t) = \mathbf{A}_0 \Pi \left([z - L_z/2]/L_z\right) e^{i(kz - \omega t + \phi_0)}$, where $\prod(x)$ is the unit box function (= 0, unless $|x| \leq 1$). The motion of an electron through each wavelet is simply motion in a plane-wave for a limited period of time $\tau_c$, equal to the fluctuation crossing time $\approx L_z |\gamma_0/v_{z0}|$. Yang [21] has obtained implicit an-alytic solutions for electron momenta $\mathbf{p}(s)$ in a plane wave with arbitrary initial momenta ($\mathbf{p}_0 = \gamma_0 m \mathbf{v}_0$) and phase $\phi_0$, in terms of the electron proper time $s(t) = \int_{t_0}^{t} \gamma(t')^{-1} dt'$. However, the relation between laboratory time (t) and proper time (s) is non-trivial: $2Rt = A_1 s + A_2 [\sin(2Rs + 2\phi_0) - \sin(2\phi_0)] + A_3 [\sin(Rs + \phi_0)]$ where $A_1 = 1 + \frac{1}{2}a^2 + (a\cos[\phi_0] + p_{x0})^2 + p_{y0}^2 + R^2$, $A_2 = a^2/4R$, $A_3 = 2a(a\cos[\phi_0] + p_{x0})/R$, $R = \gamma_0 - p_{z0}$, $a = -eE_0/\omega mc$ is the normalized vector-potential, and time is normalized using the laser frequency ($t \to \omega t$). Making the small angle approximation, $Rs \ll 1$, which corresponds to short acceleration times $t \ll \Delta t_{max}$, where $4R^2 \Delta t_{max} \approx 2(1 + \gamma_0^2 + p_{x0}^2 - 2\gamma_0 p_{z0} + p_{z0}^2) - 0.32 a p_{x0} + 0.016 a^2$, allows us to invert the expression for laboratory time and obtain: $s \simeq 2Rt/(1 + \gamma_0^2 + p_{x0}^2 - 2\gamma_0 p_{z0} + p_{z0}^2)$. This approximation, satisfied for short scale-length wavelets ($\tau_c \ll \Delta t_{max}$), allows us to obtain expressions for the change in momentum of the electron $\Delta \mathbf{p} = \mathbf{p}(\tau_c) - \mathbf{p}(0)$ when crossing a wavelet: $\Delta p_z = \{(p_{x0} + a[\cos(\phi_0) - \cos(\psi)])^2 - p_{x0}^2\}/2R$ and $\Delta p_x = a[\cos(\phi_0) - \cos(\psi)]$, where $\psi = \phi_0 + 2R^2 \tau_c/(1 + \gamma_0^2 + p_{x0}^2 - 2\gamma_0 p_{z0} + p_{z0}^2)$. These equations predict that electrons crossing a wavelet can experience a period of acceleration (deceleration) by remaining in phase with the wave during their transit, and exit the wavelet before deceleration (acceleration) occurs. By making repeated transitions of this type, electrons gain energy on average (predominantly in the forward direction), while undergoing dynamic diffusion in momentum-space.

Since most electrons move with speed $\approx c$, we can re-express the scaling in terms of plasma scale length ($L = c\tau$). This suggests increasing the plasma scale length is an equally good route to obtaining high temperatures as increasing the intensity, which may explain the high temperatures observed in experiments and PIC simulations in long scale plasmas [17, 18].

The expressions for $\Delta p_z$ and $\Delta p_x$ allow us to form a time-discrete map model in which N electrons, given random initial momenta ($-0.11 \le p_{x,z} \le 0.11$), undergo $N_t$ transitions (i): $\mathbf{p}^{i+1} \to \mathbf{p}_0^i + \Delta \mathbf{p}(\mathbf{p}_0, \phi_0)$, with the entrance phase treated as a random variable $0 \le \phi_0 \le 2\pi$. Three example energy spectra using the map are plotted in Fig. 5 for the cases $a = 0.47$ ($I = 3 \times 10^{17} W cm^{-2}$), $a = 0.66$ ($I = 6 \times 10^{17} W cm^{-2}$), and $a = 0.94$ ($I = 1.2 \times 10^{18} W cm^{-2}$), with $N = 2.5 \times 10^4$ and $N_t = 8$. The map reproduces the essential features of the 3D simulations: approximately thermal spectra, relativistic temperatures despite $a \lesssim 1$, temperature increasing with intensity. The main source of inaccuracy in this simplified picture is the assumption that electrons enter with random phase and therefore always interact with a wavelet non-adiabatically - this leads to an overestimation of the rate at which energy transfer occurs because in reality electrons do not exit a wavelet abruptly and the energy changes tend to be smaller in magnitude. Although the map model demonstrates how a plausible physical interpretation can give rise to thermal, relativistic spectra, it cannot be used in the case of long fluctuations (because it relies on $\tau_c \ll \Delta t_{max}$) and it should be used with caution at relativistic intensities because electron acceleration has been demonstrated to be chaotic in this regime [10].

We now discuss how this acceleration mechanism compares to other well known mechanisms. According to the literature, the ponderomotive mechanism occurs near the critical surface, where electrons are given a single kick by the $\mathbf{j} \times \mathbf{B}$ force in the forward direction. Unlike motion in a plane wave, the electron retains its energy when crossing the critical surface because the field decays evanescently beyond critical. However, this simplified picture does not account for the so-called ponderomotive spectrum, which contains electrons at much higher energy than the ponderomotive potential $\varphi_p$. The existence of higher energy electrons can be explained by the presence of laser turbulence, with spatial scale close to the wavelength, in the vicinity of the critical surface. In the presence of this turbulence, some electrons will undergo > 1 kicks close to the critical surface, which accounts for the electrons found at energies above $\varphi_p$. In longer scale length plasmas, most electrons are found far away from the critical surface and they have a chance to interact with the laser and plasma fields over long distances, giving rise to a host of acceleration mechanisms that explain the obsevation of very high energy tails [5, 6, 12, 14, 15] and most notably Two-Wave Chaos (TWC) [10, 11] which explains how the bulk of the spectrum can exceed ponderomotive scaling. Like laser turbulence, TWC is a rapid, highly non-linear mechanism, but it only occurs at intensities above relativistic ($I \geqslant I_{TWC} \approx 1.2 \times 10^{18} W cm^{-2}$). For intensities below $I_{TWC}$, TWC is not active, and we expect laser turbulence to dominate. We have compared acceleration in turbulent waves (as described here) and equivalent-intensity colliding waves (as in TWC) at intensities above the $I_{TWC}$ threshold and find that turbulence further enhances the characteristic TWC spectrum energy by $\approx 60\%$, and that both processes operate on similarly rapid time scales. We conclude that TWC and turbulence are approximately comparable mechanisms when $I \geqslant I_{TWC}$. The transition to TWC for $I \geqslant I_{TWC}$ is the reason for the change in the electron energy scaling with intensity near $I \approx I_{TWC}$.

In summary, we have demonstrated that spatial non-uniformities in short-pulse laser intensity profiles enables electrons to rapidly gain energy, leading to relativistic, thermal electron spectra even when the laser intensity is below the relativistic threshold.

## ACKNOWLEDGMENTS

We gratefully acknowledge useful conversations with A.Link, M.Tabak, W.L.Kruer, A.J.Kemp, G.J.Williams,

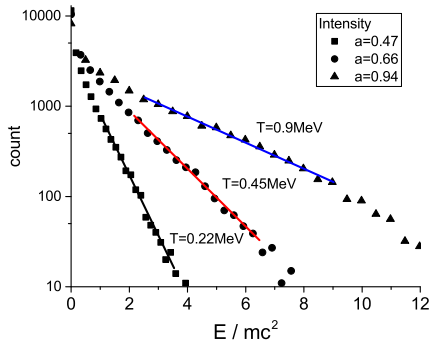

Figure 5. The electron spectrum generated by the simple map model described in the text.